%% file: dxs.tex
\documentstyle[preprint,epsf]{jpsj}
\def\figscaleA{0.9}\def\figscaleB{1.2}

\def\sf{\rm}
\input defdxm.inc
\begin{document}

\input dxs0

\sloppy
\maketitle
\pagebreak
\input dxs1
\input dxs2

\input dxs3
\input dxs4

\input dxsackn

\input dxsref
\end{document}

%% file: dxs0.tex
\def\runtitle{
Spin Excitation Spectrum of La$_{1-x}A_x$MnO$_3$
}
\title
{
Spin Excitation Spectrum of La$_{1-x}A_x$MnO$_3$
}

\def\runauthor{
Nobuo F{\sc urukawa}
}

\author{%
Nobuo F{\sc urukawa}
}

\inst{
  Institute for Solid State Physics,\\
  University of Tokyo, Roppongi 7-22-1,\\
  Minato-ku, Tokyo 106
}

\recdate
{
\hspace{4cm}
}

\abst
{
As an effective model to describe perovskite-type manganates 
(La,$A$)MnO$_3$, the double-exchange model on a cubic lattice 
is investigated.  Spin excitation spectrum of the  model in the
ground state is studied using the spin wave approximation.
Spin wave dispersion relation observed in the inelastic
neutron scattering experiment of La$_{0.7}$Pb$_{0.3}$MnO$_3$ 
is reproduced. Effective values for the electron bandwidth as 
well as Hund's coupling is estimated from the data.
}

\kword
{
perovskite-type manganate, LaMnO$_3$, double-exchange model,
metallic ferromagnetism, spin wave
}

%% file: dxs1.tex

Due to the colossal 
magnetoresistance (CMR)\cite{vonHelmolt93,Chahara93,Jin94} and other 
related phenomena, perovskite-type manganates ($R$,$A$)MnO$_3$ 
have recently been investigated intensively.
Here, A-site ions $R$ and $A$
are trivalent rare-earth and divalent alkaline-earth ions, respectively.
There exist renewed interests in these materials
 not only due to their potential
ability for application but also from the viewpoint of
strong correlation effects  in 
3$d$ transition-metal oxides which is one of the most
challenging problem in the field of the condensed matter physics.

Under appropriate hole doping,
the system becomes a metallic
ferromagnet,\cite{Jonker50a,Wollan55,Jonker56}\ {}
which is explained by the
double-exchange mechanism.\cite{Zener51,Anderson55}\ \ {}
As an effective model to describe these materials,
the double-exchange model
\bequ
  \Ham = 
  - t \sum_{<ij>,\sigma}
        \left(  c_{i\sigma}\dags c_{j\sigma} + h.c. \right)
    -\frac{J_{\rm H}}{S} \sum_i 
		\mib{S}_i \cdot \mib{\sigma}_i
    \label{HamS-finite} 
\eequ
has been proposed. Here $t$ and $\JH$ are the electron transfer energy
and Hund's coupling between electrons and localized spins, respectively.
The electron spin is represented by the Pauli matrices,
while $\mib{S}_i$ describe the localized spins.
The localized spins are considered to represent
Mn $t_{2\rm g}$ electrons with $S=3/2$ while
itinerant electrons mainly occupy the Mn $e_{\rm g}$ orbitals.

The double-exchange model correctly describes
several properties of {\LSMO}\cite{Tokura94,Urushibara95} such as 
the doping dependence of the
 Curie temperature $T_{\rm c}$\cite{Furukawa95b} and the
 universal curve of CMR
 near $T_{\rm c}$.\cite{Furukawa94,Furukawa95c}\ \ {}
However,  the double-exchange model  does not 
explain anomalous temperature dependence in the Drude weight observed
in the optical measurement of {\LSMO} at 
$ T \ll T_{\rm c}$.\cite{Okimoto95}\ \ {}
Effects of other interactions and degrees of freedom
which are present in $3d$ electron systems such as
Coulomb repulsions, lattice distortions,
orbital degeneracies and anisotropies of orbitals
 are not treated in this model.
Millis {\em et al.}\cite{Millis95x}\ {} have argued that if a polaron effect
due to strong Jahn-Teller type electron-lattice coupling  is included
in the double-exchange model, it leads to
better agreements with transport properties around $T_{\rm c}$
in doped manganates.
It is important to study
whether these effects not treated in the model
 essentially changes the properties of the model or
are able to be  taken into account  by the renormalization of
parameters $t$ and $\JH$.
In order to gain insights of  the behaviors of these materials,
it seems to be necessary to clarify to what extent
the double-exchange model correctly describes the thermodynamical
properties of perovskite manganates. 

Recently, spin wave dispersion relation of
doped manganates has been measured by the neutron inelastic scattering
experiments.
For La$_{0.7}$Sr$_{0.3}$MnO$_3$,\cite{Martin95x}\  the spin wave dispersion
relation at $T=27{\rm K}$ is given by
$\omega_q = \Delta + D_{\rm s} (q/a)^2$, where
 $D_s \simeq 12.5{\rm meV }$ is the spin stiffness
obtained from the least-squares fitting.
The spin gap probably due to anisotropies
is $\Delta \simeq 0.7{\rm meV}$,
while the lattice constant is observed as $a\simeq 3.9 {\rm \AA}$.
For
La$_{0.7}$Pb$_{0.3}$MnO$_3$,\cite{Perring95x}\  the 
observed dispersion relation fits well in the
form
\bequ
  \omega_q = \Delta +
	\Esw \frac{3 - \cos q_x - \cos q_y - \cos q_z}{6},
	\label{SW disp rel, exp Pb}
\eequ
which has been pointed out to be
reproduced by a ferromagnetic Heisenberg model
with  nearest-neighbor spin exchange couplings.
Here, the least-squares fit gives
the spin wave bandwidth  $\Esw \simeq 106 {\rm meV}$
and the gap $\Delta \sim 2.5  {\rm meV}$.

In this paper, we calculate the spin excitation spectrum 
 of the double-exchange model,
and compare the results with recent data
of the neutron inelastic scattering experiments.
We use the spin wave approximation in the ground state,
which has been introduced in the limiting case of $\JH =\infty$
by Kubo and Ohata.\cite{Kubo72}\ \ {}
We make estimates of the effective electron bandwidth and the
Hund's coupling in these materials.
The spin wave dispersion is an important measure 
to study the coherence of the electrons in the
double-exchange systems.\cite{Millis95}\ \ {}
We may obtain informations about the electronic states of doped
manganates at low temperatures as well as  their changes when
temperature is raised,
by comparing the results of the calculation with experiments.

%% file: dxs2.tex

Using the spin wave operators in the ferromagnetic state,
\bequ
   S_i{}^+ \simeq \sqrt{2S} a_i, 
\quad
   S_i{}^- \simeq \sqrt{2S} a_i\dags, 
\quad
   S_i{}^z = S - a_i\dags a_i,
		\label{defSWope}
\eequ
the Hamiltonian of the double-exchange model
is described by fermion and spin wave operators  in the form
\beqarr
  \Ham_{\rm sw} &=& \sum_{k}
		\left( 
		(\varepsilon_k - \JH) c_{k\uparrow}\dags c_{k\uparrow}
	+	(\varepsilon_k + \JH) c_{k\downarrow}\dags c_{k\downarrow}
		\right)
 		\nonumber\\
	&+ & \JH \sqrt{\frac{2}{SN}} \sum_{qk}
		\left( a_q\dags c_{k\uparrow}\dags c_{k+q\downarrow}
          +        a_q  c_{k+q\downarrow}\dags c_{k\uparrow}
		\right)
		\nonumber\\
	&+ &  \frac{\JH}{SN} \sum_{kq_1q_2\sigma}
		\sigma a_{q_1}\dags a_{q_2}
			c_{k-q_1\sigma}\dags c_{k-q_2\sigma}.
\eeqarr
We  restrict ourselves to 
the lowest order terms of the $1/S$ expansion at $T=0$.
We consider the case of $\JH$ being finite but sufficiently large
so that electrons are also completely polarized, {\em i.e.} 
$n_\uparrow=n$ and $n_\downarrow=0$, at $T=0$.
The electron concentration is described as $n=1-x$.
We assume a simple cubic lattice with nearest-neighbor
electron hoppings so that  we have
\bequ
  \varepsilon_k = -2t \left(
	\cos k_x + \cos k_y + \cos k_z
	\right).
		\label{def DispRelFermi cubic}
\eequ
For perovskite manganates, estimates of 
the electron bandwidth and the on-site Hund's coupling
being a few {\rm eV} has been made by the 
first-principle calculations.\cite{Hamada95}\ \ {}
However, we consider that  $t$ and $\JH$ in this model
are the effective parameters which could be strongly renormalized from the
bare value due to other interactions present in the real systems.
 These parameters
 should be determined from  comparisons with experiments.

\begin{figure}
\hfil{\def\epsfsize#1#2{\figscaleB#1}\epsfbox{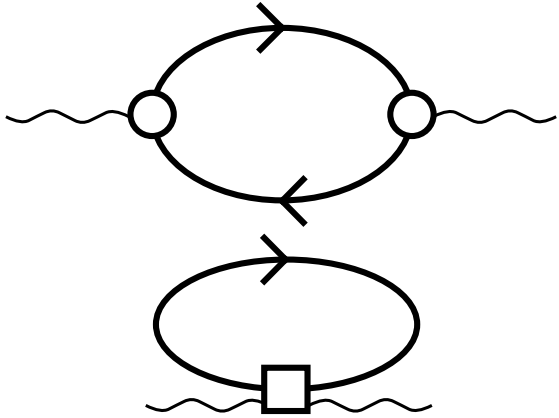}}\hfil
\caption{Spin wave self energy in the lowest order of the $1/S$ expansion. 
Green's functions for electrons
and spin wave are given by bold curves and wavy lines, respectively.
Vertices for $xy$ and $z$ couplings are represented by circles and 
a square, respectively.}
\label{FigSWPi}
\end{figure}

\begin{figure}
\hfil{\def\epsfsize#1#2{\figscaleA#1}\epsfbox{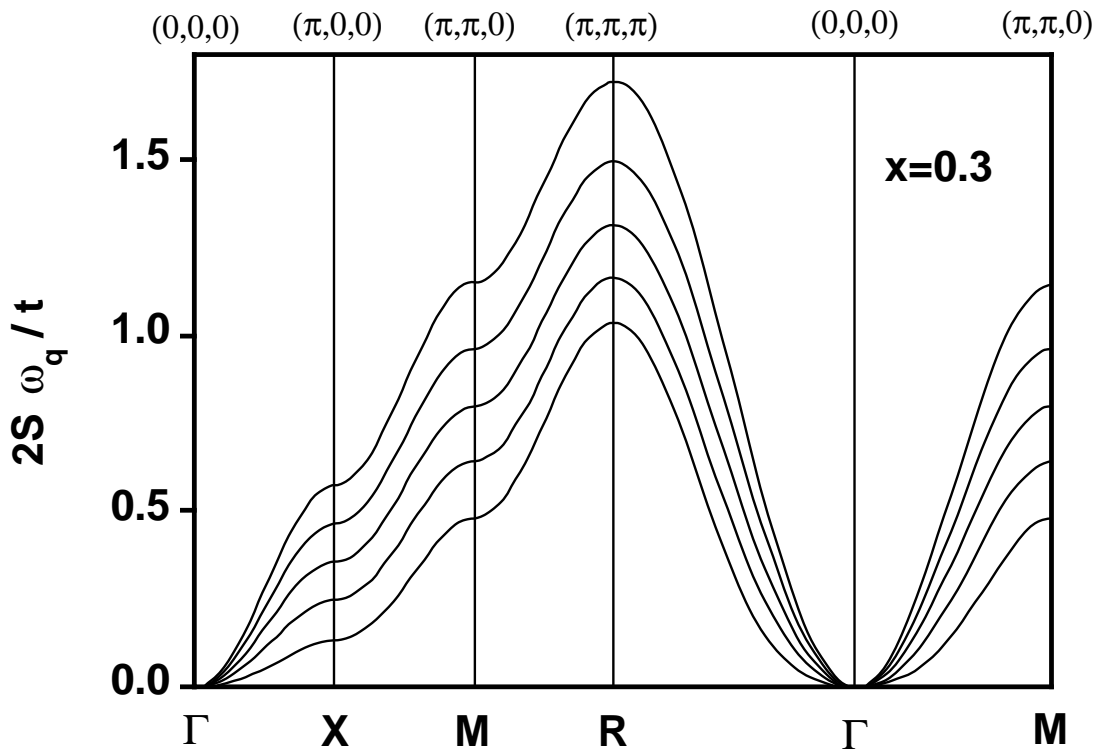}}\hfil
\caption{Spin wave dispersion relation at $x=0.3$.
Curves are for $\JH/t=\infty$, $24$, $12$ and $6$ from top to bottom.}
\label{FigSpinDisp}
\end{figure}

The spin wave self-energy in the lowest order of $1/S$ expansion,
which is schematically illustrated in Fig.~\ref{FigSWPi},
is given by
\beqarr
  \Pi(q,\omega) &=& \frac{1}{SN}
    \sum_k \left(f_{k\uparrow} - f_{k+q\downarrow} \right)
			\nonumber\\
	& \times&
	\left(\JH + \frac{2\JH^2}
		{ 
	  \omega + \varepsilon_k - \varepsilon_{k+q} - 2\JH } 
	\right),
    \label{fmlaPi}
\eeqarr
where $f_{k\sigma}$ is the Fermi distribution function.
We have
$f_{k\downarrow} = 0$ from the assumption.
The spin wave dispersion relation $\omega_q$ is obtained self-consistently
as a solution of
the equation $  \omega_q = \Pi(q,\omega_q)$.
Since $\Pi \propto 1/S$, the lowest order $1/S$ expansion gives
$\omega_q = \Pi(q,0)$. Therefore, the spin wave dispersion
is described as
\bequ
    \omega_q = 
 \frac{1}{2S} \frac{1}{N}
    \sum_k f_{k\uparrow}
	  \frac{\JH(  \varepsilon_{k+q} - \varepsilon_{k} )}
		{ \JH + (\varepsilon_{k+q} - \varepsilon_{k} )/2 } .
	\label{SWdispreldef}
\eequ
In Fig.~\ref{FigSpinDisp}, we show the spin wave dispersion relation
at $x=0.3$ for various values of $\JH/t$.
As the value of  $\JH$ becomes comparable with the electron bandwidth,
we see the softening of the spin wave dispersion since the
effective coupling between spins become weak.

At $\JH \to\infty$, we have 
\beqarr
 \omega_q &\simeq&  \frac{1}{2SN}\sum_k   
			( \varepsilon_{k+q} - \varepsilon_{k})
       f_{k\uparrow}
	\label{defSWDRJinf} 	\\
      &=& 
    \Esw \frac{3 - \cos q_x - \cos q_y - \cos q_z }{6},
	\label{SW, allBZ, nn}
\eeqarr
where $\Esw $ is the spin wave bandwidth given by
\bequ
  \Esw = \frac{6t}{SN} \sum_k f_{k\uparrow} \cos k_x.
\eequ
The dispersion relation (\ref{SW, allBZ, nn}) is the same
as those in a ferromagnetic Heisenberg model with nearest-neighbor
spin exchanges.

The above correspondence can be understood as follows.
We consider a perfectly spin polarized state at $T=0$ and then
twist a spin at site $i_0$.  In the case of the strong coupling limit
$\JH \gg t$ where electrons with spins anti-parallel to the localized spins
of the same cite are disfavored, 
the electron at site $i_0$ is localized because
it has different spin orientation to the localized spins in neighboring sites.
Therefore, in this limit the effective spin-spin interaction is short
ranged. As $\JH/t$ increases, electrons become more localized so
the range of effective interaction becomes shorter.
It is in a sharp contrast to the weak coupling limit
where the RKKY long range
interaction with power law decay is mediated by the coherent motion
of electrons.
As long as we restrict ourselves to the
single magnon excited states at $T=0$,
 the double-exchange model in the strong coupling limit
is mapped to the Heisenberg model with short-range interactions.
%

%% file: dxs3.tex

%
\begin{figure}
\hfil{\def\epsfsize#1#2{\figscaleA#1}\epsfbox{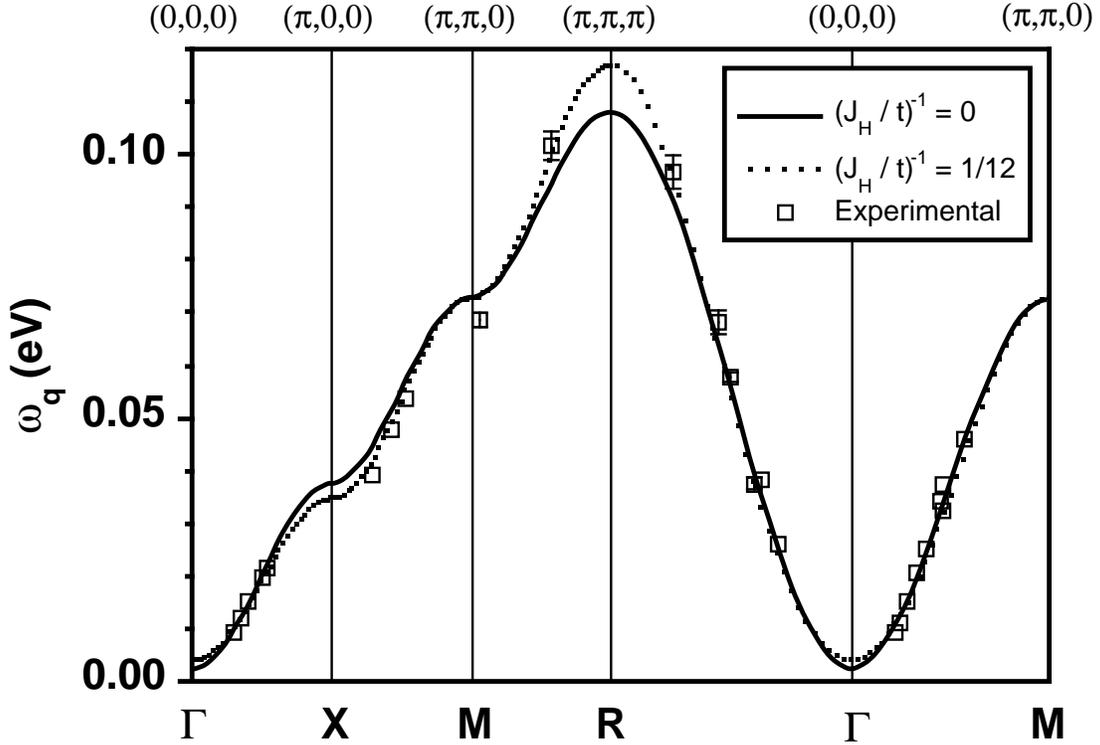}}\hfil
\caption{Spin wave dispersion relation of \LaPbMnO (squares).
Curves are calculated by fitting data at $\JH/t=12$ (dotted curve)
 and $\JH/t = \infty$ (solid curve).}
\label{FigNeutron}
\end{figure}

Now we compare the results with experiments.
In Fig.~\ref{FigNeutron}, we plot the data of the neutron
inelastic scattering experiment\cite{Perring95x} 
together with  the fitting
curves in the form
\bequ
  \omega_{\rm fit}(q) 
   = \Delta + \omega_q,
\eequ
where $\omega_q$ is obtained at $x=0.3$ as
a function of $\JH$ and $t$ from eq.~(\ref{SWdispreldef}).
It has been pointed out that the
dispersion relation in La$_{0.7}$Pb$_{0.3}$MnO$_3$ 
fits well in the cosine-band form
with  $\Esw = 0.1055{\rm eV}$ and 
$\Delta= 0.0025{\rm eV}$.\cite{Perring95x}\ {}
As shown previously, the double-exchange model at $\JH/t \to \infty$
 gives the same dispersion relation, with the electron hopping
energy  $t=0.18{\rm eV}$.
At $\JH/t = 12$, which is in the intermediate coupling region,
we have a better fit to the data at
$t=0.26{\rm eV}$ and $\Delta =0.004{\rm eV}$.
Thus we see that the dispersion relation of the 
spin wave obtained experimentally 
is reproduced by the double-exchange model in
a realistic parameter region $\JH / t \simge 12$ and $t\sim 0.2{\rm eV}$.

\begin{figure}
\hfil{\def\epsfsize#1#2{\figscaleA#1}\epsfbox{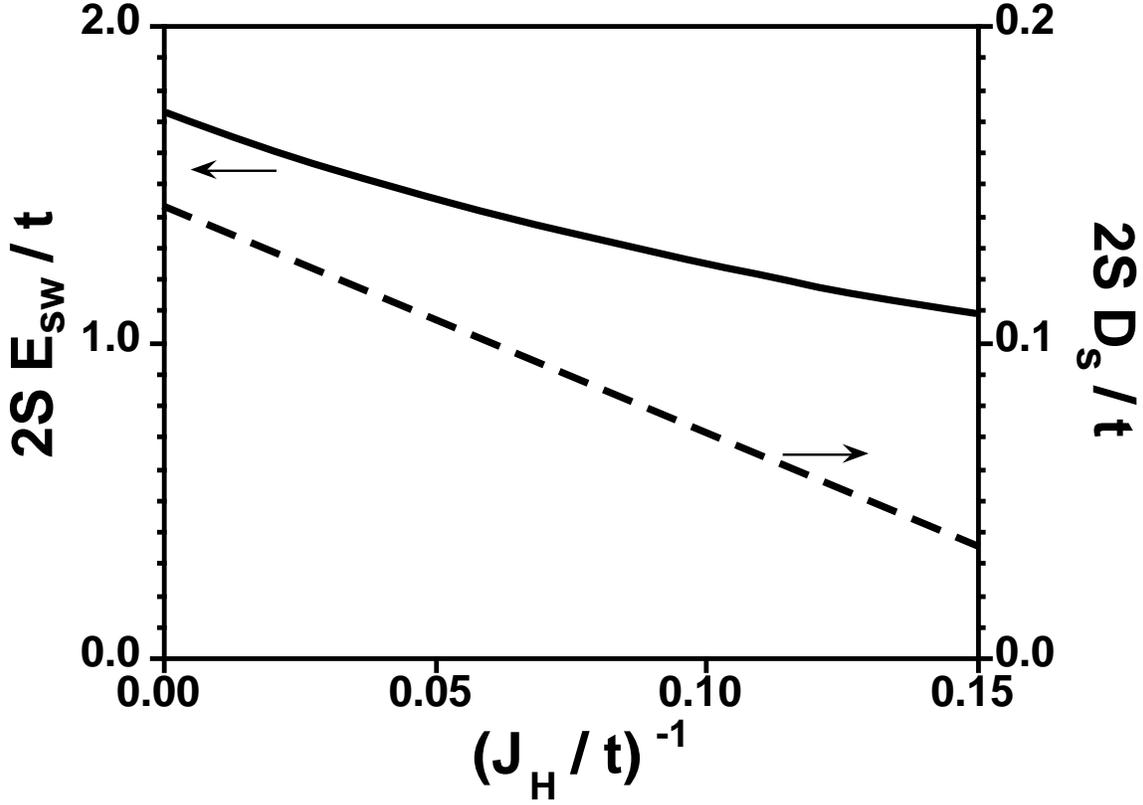}}\hfil
\caption{Magnon bandwidth $\Esw$ (solid curve)
 and the spin stiffness $D_{\rm s}$ (dashed line) at $x=0.3$.}
\label{FigEswD}
\end{figure}

For {\LaSrMnO}, the hopping energy $t$ 
is estimated as follows.
The spin stiffness of {\LaSrMnO} is obtained experimentally as
$D_{\rm s} \simeq 12.5{\rm meV}$.
From the model calculation,
we expand eq.~(\ref{SWdispreldef}) 
at $q\to0$ and obtain the spin stiffness.
On a cubic lattice with the electron dispersion relation
(\ref{def DispRelFermi cubic}), we have 
$   \omega_q = D_{\rm s} q^2 $
where the spin stiffness constant $D_{\rm s}$ is given by
\bequ
  D_{\rm s} = \frac{t}{2S} \frac1N\sum_k f_{k\uparrow}
  \left(  \cos k_x - \frac{2t}{\JH} \sin^2 k_x \right).
\eequ
In Fig.~\ref{FigEswD}, we show 
 the spin stiffness $D_{\rm s}$ at $x=0.3$ as a function of $\JH/t$,
together with the spin wave bandwidth
$\Esw \equiv \omega_{q=Q} - \omega_{q=0}$ .
At $S=3/2$, we have $D_{\rm s} < 0.04t$ for finite $\JH/t$.
Then we  have the estimated value as $t \simge 0.3 {\rm eV}$.

Let us now investigate the 
effective electron hopping energy at $T \sim T_{\rm c}$.
From the above effective parameters for {\LaPbMnO}, we 
calculate the Curie temperature 
 and  make a comparison with experiments.
In order to obtain $T_{\rm c}$ in the double-exchange model,
we apply the infinite-dimensional approach.
For details of the calculation, readers are referred to
ref.~\citen{Furukawa95b}.
Here we simply use the result for the semicircular density of states
where the bandwidth is taken to be equal to the value
on the cubic lattice,  $W \equiv 6t$.
In the case $\JH/t = 12$ or $\JH/W=2$, 
 we obtain $T_{\rm c}^\infty = 0.146t$ at $x=0.3$.
Then, the previous value $t\sim 0.26{\rm eV}$ obtained from the fitting
 gives $T_{\rm c}^{\infty} \sim 440{\rm K}$.
Experimentally,  we observe $T_{\rm c}^{\rm exp} = 355{\rm K}$ in {\LaPbMnO}. 
Since  spatial fluctuations which are absent in infinite-dimensional
approaches reduce $T_{\rm c}$,
we consider that
the parameters determined at $T=0$ may also 
describes $T_{\rm c}$ for {\LaPbMnO} in a consistent way.
Thus, the double-exchange model with a fixed parameters $t$ and $\JH$
simultaneously accounts for the spin wave dispersion relation
at the low temperature region
 and the Curie temperature $ T_{\rm c}$.

There exists no substantial change in the electronic hopping energy
at $T \sim T_{\rm c}$ and $T \sim 0$ in these perovskite manganates.
The value of $t$ estimated above is comparable with those 
obtained from the first-principle calculation. Therefore, 
 in these compounds, we consider
that  the renormalizations of electron coherence 
due to dynamical lattice distortions and other elementary excitations
are not so large and do not have substantial temperature dependences.
As long as  {\LaPbMnO} is concerned,
the double-exchange model alone seems to describe the
spin excitation properties adequately.

It has been known that Sr- and Pb-doped lanthanum manganates have
wider bandwidth compared to other compounds in this family 
of materials.\cite{Jonker50a}\ \ {}
At $x=0.3$, the above manganates are far from the phase boundaries of
insulator-metal transition or structural transition
and are metallic even in the paramagnetic phase.
Since the double-exchange model reproduces the thermodynamical
properties of these compounds,
we consider that  the effect of coherent motions of
electrons dominantly determine the thermodynamical properties
of these relatively wide-banded compounds in the metallic region, indeed.
Effects of Coulomb interactions and Jahn-Teller type
distortions may cause renormalization
of electron hopping energy and Hund's coupling,
 but they seem to be irrelevant to the low energy properties.
Further systematic studies of the other bandwidth-controlled manganates
will show us how and in what condition these residual interactions
become relevant.

%% file: dxs4.tex

We note here the relation between the present calculation and the
approach by Kubo and Ohata.\cite{Kubo72}\ \ {}
They have considered the double-exchange model in the
strong Hund's coupling limit $\JH/t=\infty$.
The Hamiltonian is given by
\bequ
  \Ham' =
	-t \sum_{\stackrel{<ij>} { \sigma\sigma'}}
	\left(
    c_{i\sigma}\dags   (\hat P_i \hat P_j)_{\sigma\sigma'}c_{j\sigma'}
		+ h.c.
	\right),
      \label{HamPJ}
\eequ
where $\hat P_i = \hat P(\mib{S}_i)$ 
is the projection operator to the electronic state with a spin
parallel to the localized spin at $i$-th site $\mib{S}_i$.
At the first and the second order expansion of $1/S$,
the present approach in the limit $\JH/t\to\infty$
 and the spin wave approximation for
the projected Hamiltonian (\ref{HamPJ}) give equal results
for the spin wave dispersion
in the single magnon excited state at $T=0$.
Although the interaction vertices are proportional to $\JH$, 
the spin wave self energy at $\JH/t \to\infty$ does not show singularities 
but converges to those obtained at $\JH/t=\infty$,
at least up to the second order of $1/S$ expansions.
Therefore, in the low order $1/S$ expansions,
 the present  approach for finite $\JH/t$ seems to be valid 
even in the large Hund's coupling region.
Details on higher order expansions with respect to $1/S$
 will be reported elsewhere.


To summarize, we have calculated the spin wave dispersion relation
of the double-exchange model using the spin wave approximation.
Comparison with the neutron scattering
experiment data 
in {\LaSrMnO} and {\LaPbMnO}
has been performed. 
Spin wave dispersion relation throughout the Brillouin zone 
in {\LaPbMnO} is reproduced qualitatively.
Values of the electron bandwidth as well as Hund's coupling
 is estimated from the spin wave bandwidth and the spin stiffness.
The double-exchange model consistently describes the
spin wave dispersion relation and the Curie temperature
of {\LaPbMnO}.

%% file: dxsackn.tex
\section*{Acknowledgments}

The author would like to thank Y. Endoh and Y. Tokura
for fruitful suggestions.
This work was supported by a Grant-in-Aid for 
Encouragement of Young Scientists
from the Ministry of Education, Science and Culture.